# Comment on "Influence of non-conservative optical forces on the dynamics of optically trapped colloidal spheres: The fountain of probability", cond-mat.soft 0804.0730v1

Rongxin Huang, Pinyu Wu, and Ernst-Ludwig Florin

*Center for Nonlinear Dynamics, University of Texas, Austin, Texas 78712, USA*

In a recent letter, Y. Roichman *et al.*, [1] described experiments they performed in order to demonstrate that an optically trapped colloid particle reaches a non-equilibrium steady state within a static optical trap with its probability density showing steady toroidal currents, a phenomenon they call the "fountain of probability". They concluded that a 2.2 μm colloid silica sphere undergoes clockwise circulations in the optical trap due to non-conservative radiation forces. Assuming that the scattering force has its maximum on the optical axis and decays away from it, they present a picture in which the particle is pushed forward along the optical axis and returns back at the periphery of the trap. The authors support their findings by Brownian dynamics simulations performed with the experimental parameters and find that their description of the scattering force is sufficiently detailed for a quantitative comparison with the experimental data. However, following the authors' "measure of a trajectory's circulation" ($\Omega(t)$) and of the "accumulation of circulation" ($\chi(t)$), we are not able to reproduce the observed circulation in Brownian dynamics simulations with the given experimental parameters. Moreover, a simple geometrical optics calculation of the scattering forces acting on a 2.2μm particle shows that the assumption of a Gaussian profile for the strength of the scattering force is not appropriate and, therefore, the agreement of Brownian dynamics simulations and experimental data in the manuscript is surprising. Considering both the inconclusive description of the circulation and the inaccurate description of the scattering force, the manuscript does not present an experimental verification of the action of non-conservative forces in an optical trap.

**The measure of circulation**

The authors introduce a measure of the trajectory's circulation (Eq. (3) and (4)), which they define as the total area swept by the particle normalized to the area of the trap's cross section for a given time interval. To measure the trend of circulation in the trap, they integrated the area over a time series. An increasing trend in the integral ($\chi(t)$) represents a clockwise motion whereas a negative trend represents a retrograde motion. To confirm the validity of this measure of circulation, we performed conventional Brownian dynamics simulations with the parameters given in the manuscript [2]. We chose short time steps of $1 \times 10^{-4}$ second. Figure 1 (a) shows the circulation calculated for a particle in a three-dimensional harmonic potential with force constants taken from the manuscript and points taken 1/30 second apart. As expected for a particle moving in a harmonic potential, there is an accumulated circulation fluctuating around zero (green trace), but accumulated circulations with a positive or a negative trend (red and blue traces) are also observed with equal probability in different runs of the simulation. In the next set of simulations, we included a scattering force as described in the manuscript $f_1 \exp(-\frac{r^2}{2\sigma^2})\hat{z}$ while keeping all other parameters as described before. Figure 1 (b) shows three examples of accumulated circulations. As for the case of a purely harmonic potential, positive as well as negative trends are observed with equal probability. Thus, we conclude that the described measure for the accumulated circulation is insufficient in the parameter range described in the manuscript.

**Scattering force profile for a 2.2 μm silica sphere at 512 nm**

The authors assume that the scattering force field follows a Gaussian profile with its maximum on the optical axis and a width on the order of the particle diameter. With this assumption, they postulate the particle must be pushed forward along the optical axis and return back when it is away from optical axis where the scattering force is weaker, resulting in a clockwise circulation. However, given the particle diameter of 2.2 μm and the trapping laser wavelength of 512 nm, the scattering force can be well-approximated by the ray optics model [3, 4]. The calculated scattering force (Figure 2) has a minimum on the optical axis and increases drastically for larger displacements, in stark contrast to the assumption made in the manuscript. Therefore, if there existed a net circulation of a particle in the optical trap and it could be quantified, the sign of the circulation should be the opposite of the reported circulation.

In summary, we conclude that the presented data in the manuscript do not provide sufficient evidence for a circulation driven by the scattering force in a static optical trap. An improved method might be necessary for measuring the effect of non-conservative forces on the trapped particle.

## References

[1] Yohai Roichman, Bo Sun, Allan Stolarski, David G. Grier, cond-mat.soft/0804.0730v1 (2008).

[2] A.C. Branka, D.M. Heyes, Phys. Rev. E **58**, 2611 (1998).

[3] A. Ashkin, Biophys. J. **61**, 569 (1992).

[4] J.Y. Walz, Appl. Opt. **38**, 5319 (1999).

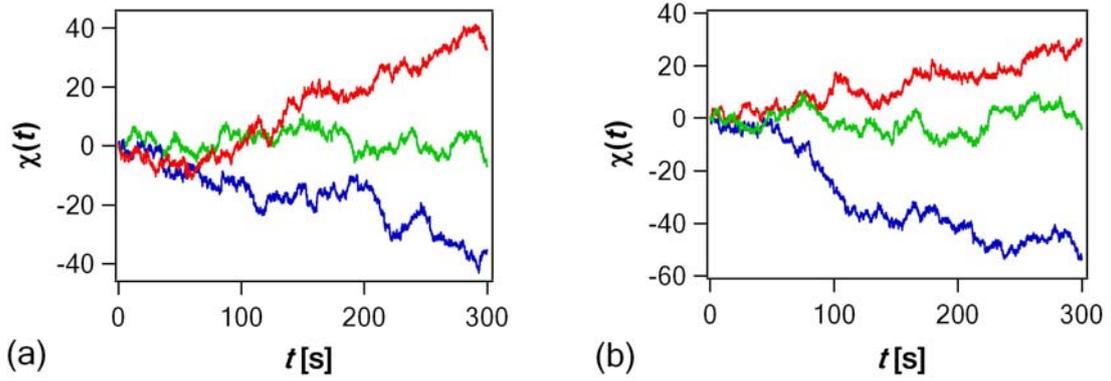

**Figure 1:** (a) Three typical circulation traces for a trapped particle moving within a three-dimensional harmonic potential. The conventional Brownian dynamics simulation has been performed in a three-dimensional harmonic potential with $k_x$=0.467 pN/μm, $k_y$=0.4 pN/μm and $k_z$=0.08 pN/μm with time steps of $1*10^{-4}$ second for $3*10^6$ points per trace. The radius of the particle was assumed to be $a$ =1.1 μm, the same as in the described experiments. The circulation was calculated using Eq. (3) and (4) in [1] for points with 1/30 second apart. (b) Three typical circulation traces for a particle moving in the same harmonic force field as described above plus a scattering force field defined as $f_1 \exp(-\frac{r^2}{2\sigma^2})\hat{z}$ with $\sigma = a = 1.1$ μm and $f_1 = 0.1*k_z*\sigma$.

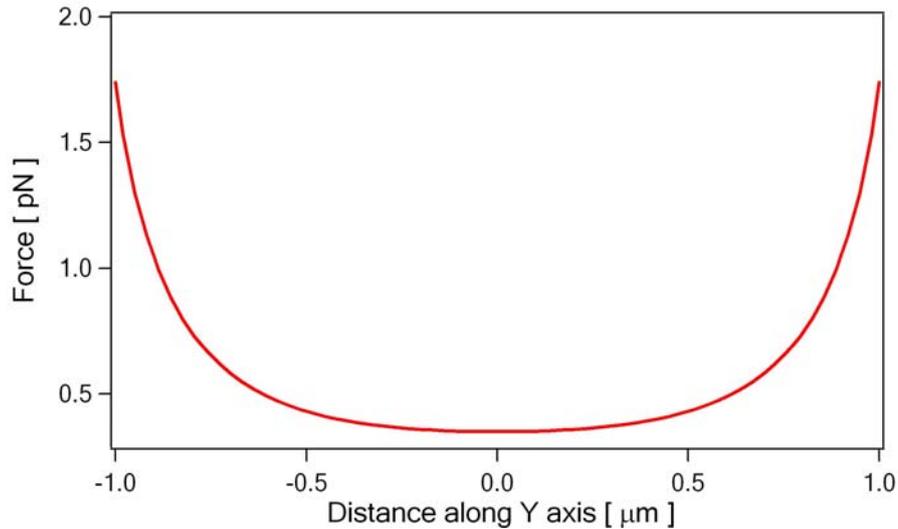

**Figure 2:** Scattering forces acting on a particle when the center of the particle is displaced laterally along the y-axis. A particle radius of 1.1 μm and a refractive index of 1.50 were assumed. The refractive index of the medium was assumed to be 1.33. The objective lens has a numerical aperture of 1.4 and a back aperture with a radius of 4 millimeters. Laser intensity distribution has been assumed to be Gaussian with a beam radius equal to the radius of the back aperture of the objective lens. The wavelength of the laser was assumed to be 512 nm and the total laser power at the focal plane was assumed to be 7.5 mW. Although the numerical aperture used in holographic tweezers for an individual trap might be smaller than 1.4 and other parameters used in the calculation such as the radius of the back aperture of the objective lens, the laser power at the focal plane, etc. might not be the same as in the described experiments, the overall profile of the scattering force along y-axis will not change.